\documentclass[10pt,twocolumn,journal]{IEEEtran}
\usepackage{times}
\usepackage{amsbsy}
\usepackage{latexsym}
\usepackage{color}
\usepackage{amssymb}
\usepackage{amsmath}
\usepackage[ansinew]{inputenc}
\usepackage[dvips ]{graphicx}
\input epsf
\usepackage{epic,eepic}
\usepackage{url}
\usepackage{caption}
\title{\huge Multiple-Candidate Successive Interference
Cancellation with Widely-Linear Processing for MAI and Jamming
Suppression in DS-CDMA Systems }
\author{Jianwei Yang$^1$  Rodrigo C. de Lamare$^2$ \\
$^1$School of Mechatronics Engineering and Automation, National University of Defense Technology, Changsha, 410073, People's Republic of China\\
$^2$CETUC/PUC-Rio, Brazil and Department of Electronics, University of York, YO10 5DD, UK \\
E-mail: zhsh94@hotmail.com }

\begin{document}
\maketitle
\begin{abstract}
In this paper, we propose a widely-linear (WL) receiver structure
for multiple access interference (MAI) and {jamming signal (JS)}
suppression in direct-sequence code-division multiple-access
(DS-CDMA) systems. A vector space projection (VSP) scheme is also
considered to cancel the {JS} before detecting the desired signals.
We develop a novel multiple-candidate successive interference
cancellation (MC-SIC) scheme which processes two consecutive user
symbols at one time to process the unreliable estimates and a number
of selected points serve as the feedback candidates for interference
cancellation, which is effective for alleviating the effect of error
propagation in the SIC algorithm. Widely-linear signal processing is
then used to enhance the performance of the receiver in non-circular
modulation scheme. By bringing together the techniques mentioned
above, a novel interference suppression scheme is proposed which
combines the widely-linear multiple-candidate SIC (WL-MC-SIC)
minimum mean-squared error (MMSE) algorithm with the VSP scheme to
suppress MAI and {JS} simultaneously. Simulations for binary phase
shift keying (BPSK) modulation scenarios show that the proposed
structure achieves a better MAI suppression performance compared
with previously reported SIC MMSE receivers at lower complexity and
a superior {JS} suppression performance.
\end{abstract}


\section{introduction}
Direct-sequence code-division multiple-access (DS-CDMA) systems are
one of the most successful multiple access technologies for wireless
communication systems. Such services include third generation
cellular telephone, indoor wireless networks and satellite
communication systems. Multiple access interference (MAI), arising
from the nonorthogonality between the signature sequences and
jamming signals is a significant limiting factor to the performance
of DS-CDMA systems. To address this problem, the optimum multiuser
detector (MUD) and several suboptimum MUDs were introduced
{\cite{Verdu}}. Optimal multiuser detection has an exponential
computational complexity and is therefore impractical. Several
low-complexity multiuser detectors including the linear
decorrelator, the linear minimum mean-squared error (MMSE), the
successive interference cancellation (SIC) \cite{Holtzman}, and the
parallel interference cancellation (PIC) have been proposed
\cite{Divsalar}.

The SIC detector regenerates and cancels the signals of other users
before data detection of the desired user. The potential of SIC to
alleviate the near–far problem comes from its property of removing
stronger users before detecting weaker users. Specific forms of the
SIC are closely related to approximations of the optimum maximum
likelihood (ML) detector \cite{{Tan},{Nelson}}, as well as iterative
techniques for solving linear equations \cite{Schotten}. There are
several variants of the SIC algorithm, which have been investigated
in last decade or so \cite{Shynk}-\cite{fa}. The performance of the
SIC algorithm relies heavily on the accuracy of the symbol estimate
and is subject to error propagation effects when the estimated
symbol is not accurate. Methods including soft or linear
interference cancellation and partial interference cancellation are
proposed to mitigate this error propagation \cite{Blostein}. A
multiple feedback SIC (MF-SIC) algorithm with shadow area
constraints (SAC) strategy for detection of multiple data streams
has been introduced in \cite{Peng1}. The MF selection algorithm
searches several constellation points rather than one and chooses
the most appropriate constellation symbol as the decision. A joint
successive interference cancellation technique (JSIC) has been
introduced in \cite{Singer}. The key idea behind JSIC is to exploit
the structural properties of the sub-constellation formed by the
signals of two consecutive users in an ordered set to gain an
improvement in the detector performance.

In wireless communication systems, most interference suppression or
parameter estimation techniques are based on linear signal
processing \cite{Lamare,sheng}. However, when a noncircular
modulation is applied, e.g. binary phase shift keying (BPSK), linear
estimation of an improper real-valued signal from complex data
appears complex, which in a statistical signal processing sense, is
not optimal. It has been shown in \cite{Chevalier} that by
exploiting the improper nature of the received signal, the
estimation performance can be significantly improved. Therefore, the
resulting widely linear (WL) estimate has gained great popularity
for systems using non-circular modulation schemes
\cite{Schreier,Buzzi,song}.

The jamming signal (JS) is another form of interference which has a
huge influence on the performance of DS-CDMA systems. The notch
filter is used to cancel the  {JS} and an estimate of the
interference parameters is required before the interference
cancellation \cite{Milstein}. A generalized approach for the  {JS}
suppression in PN spread-spectrum communications using open-loop
adaptive excision filtering is introduced \cite{Alan}. The algorithm
has a tradeoff between interference removal and the amount of
self-noise generated from the induced correlation across the PN chip
sequence due to the filtering procedure. A new transversal filter
structure is used before correlation to improve the performance of
DS-CDMA systems \cite{Panay}. Several subspace techniques have been
developed to exploit the low-rank structure of the interference in .
The eigenspace-based interference canceller has been proposed in
\cite{Zhuang} and  to cancel the interference through constructing
the proposed estimate-and-subtract interference cancellation
beamformer.

The goal of this work is to develop an interference suppression
strategy for DS-CDMA systems that operates in the presence of
non-circular data,  {JS} and MAI. To this end, we bring together a
novel SIC algorithm, WL processing and a  {JS} cancellation scheme.
Inspired by the error propagation mitigation in the MF-SIC and JSIC
algorithms, we propose a novel Multiple-Candidate SIC (MC-SIC)
scheme which processes two consecutive user symbols at one time when
the current symbol decision is not reliable and also exploits the
constellation knowledge in the generation of candidates for
detection. In addition to this, we combine WL processing with the
MC-SIC scheme and propose a widely linear MC-SIC (WL-MC-SIC)
algorithm, which aims to deal with the performance degradation due
to error propagation in SIC and linear signal processing on the
noncircular signal. We also devise a technique to cancel JS that is
incorporated into the proposed WL-MC-SIC algorithm, which relies on
a vector space projection (VSP) method. The interference
cancellation operator is constructed to project the desired signal
onto the complement of the JS subspace.

The main contributions of this paper are summarized as follows:
\begin{itemize}
\item  A MC-SIC scheme based on the MMSE criterion is proposed.

\item  A novel widely linear MC-SIC (WL-MC-SIC) MMSE algorithm is devised, which combines WL processing and a MC-SIC MMSE scheme to improve the
detector’s performance under noncircular modulation scheme and alleviate the effect of error propagation in the traditional SIC algorithm.

\item A novel interference mitigation scheme is proposed which combines the VSP algorithm with the WL-MC-SIC MMSE algorithm to jointly suppress the MAI and  {JS}.

\item  The performance of the proposed algorithm for MAI and  {JS} suppression is compared with other interference suppression schemes.
\end{itemize}

The organization of the paper is as follows. The system model is
given in Section II. Section III introduces the proposed WL-MC-SIC
algorithm and VSP. In Section IV the complexity of the WL-MC-SIC
algorithm is analyzed. Section V presents the simulation results and
a comparison between the proposed WL-MC-SIC algorithm with the VSP
method and previously reported algorithms.

Notation: In this paper, scalar quantities are denoted with italic typeface. Lowercase boldface quantities denote vectors and uppercase boldface quantities denote matrices. The operations of transposition, complex conjugation, and conjugate transposition are denoted by $\{\cdot\}^{T}$, $\{\cdot\}^{\ast}$ and $\{\cdot\}^{H}$, respectively. The symbol $\textrm{E}\lbrack\cdot\rbrack$ denotes the expected value of a random quantity, the operator $\mathcal{R}\{\cdot\}$ selects the real part of the argument, the operator $\mathcal{J}\{\cdot\}$ selects the imaginary part of the argument, $\mathcal{L}[\boldsymbol{M}]$ denotes a linear subspace spanned by the columns of the matrix A and the operator $|\cdot|$
denotes the absolute value of the argument.

\section{system model}
Let us consider a synchronous DS-CDMA system with K active users signaling through an additive Gaussian noise channel. The received baseband signal during one symbol interval in such channel can be modeled as
\begin{equation}
\begin{split}
r(t) & = \sum_{k=1}^{K}  {A_{k}}  {b}_{k} s_{k}(t) +
j(t)+n(t),~~~t\in[0,T],\label{recsignal_analogy}
\end{split}
\end{equation}
where $j(t)$ and $n(t)$ represent the jamming signal and the ambient channel noise respectively; $T$ is the symbol interval; $b_{k}\in\{\pm{1}\}$ is the BPSK symbol for the user $k$ with amplitude $A_{k}$, and $s_k(t)$ is the spreading sequence
waveform of the $k$-th user.

After chip matched filtering and time synchronization, sampling with rate ${L}/{T}$, where $L$ is equal to the spreading factor and appropriate normalization, the vector $\boldsymbol{r}(i)$ containing the $L$ samples received in the interval $[iT;(i+1)T),~i\in{\mathbb{Z}}$ can be expressed as
\begin{equation}
\begin{split}
{\boldsymbol r}[i] & = \sum_{k=1}^{K} A_{k}{b}_{k}[i]
{\boldsymbol s}_{k}[i] + {\boldsymbol{j}[i]} + {\boldsymbol n}[i], \label{recsignal_vector}
\end{split}
\end{equation}
where $\boldsymbol{s}_{k}$ is the spreading sequence vector of the $k$th user. The quantity $\boldsymbol{j}[i]$ is the jamming interference
vector, $\boldsymbol{n}[i]$ is a zero-mean additive white Gaussian noise (AWGN) sample vector with $\boldsymbol{E\{\boldsymbol{n}[i]\boldsymbol{n}^{H}[i]\}}={\sigma}^{2}_n\boldsymbol{I}_{L}$, ${\sigma}^{2}_n\triangleq{N_{0}}$,
where $N_{0}$ is the single-sided power spectral density, $\boldsymbol{I}_{L}$ is the $L\times{L}$ identity matrix. We use the definition $\boldsymbol{p}_{k}\triangleq{A_{k}\boldsymbol{s}_{k}}$, $\boldsymbol{P}\triangleq{[\boldsymbol{p}_{1} \cdots {\boldsymbol{p}_{k}}]}$ and $\boldsymbol{b}{[i]}\triangleq{[b_{1}[i] b_{2}[i] \cdots {b_{K}[i]}]^{T}}$. Eq.{(\ref{recsignal_vector})} can be equivalently expressed as
\begin{equation}
\begin{split}
{\boldsymbol r}[i] & = {\boldsymbol{P}}\boldsymbol{b}[i]+\boldsymbol{j}[i]+\boldsymbol{n}[i]. \label{rec_vector}
\end{split}
\end{equation}
We assume the transmitted symbol sequences of different users are mutually and statistically independent. The spreading sequences are linearly
independent and normalized to $\boldsymbol{s}^{H}\boldsymbol{s}=1$,~$1\leq{k}\leq{K}$.

The jamming signal is modeled as a sinusoidal signal (tone) or an
autoregressive (AR) signal. The jamming signal can also be digital
with a data rate much lower than the spread spectrum chip rate. The
tone interference is commonly used in the  {JS} analysis
\cite{Poor1} and we use this type of interference as the
 {JS}, which can be expressed as
\begin{equation}
\begin{split}
{\boldsymbol j}(i) & = \sum_{l=1}^{m}\sqrt{P_{i}} \textrm{e}^{\textrm{j}(2 {\pi} f_{i} l +\varphi_{l})}, \label{jamming_signal}
\end{split}
\end{equation}
where $P_{i}$ and $f_{i}$ are the power and the normalized frequency
of the $i$th tone interference and  {$\varphi_l$} are independent
random phases uniformly distributed on $(0,2\pi)$. The quantity $m$
denotes the number of components of the tone interference.

In this paper, we detect the users according to their received power level arranged in descending order; the strongest user is detected first.
We assume that the user $k$ is the desired user and estimate the user $k$ after removing $k-1$ users from the received signal. For our
implementation of the SIC schemes, we assume perfect knowledge of the signal amplitudes and the spreading codes.

\section{Proposed Widely Linear MC-SIC MMSE Detector Design}
In this section, we firstly introduce the scheme of the VSP
interference canceller \cite{{Zhuang}} to suppress the  {JS} before
estimating the desired users. Secondly, we review the linear MMSE
detector based on widely linear signal processing. Then we describe
the MC-SIC MMSE detector design based on the constellation
constrains and multiple-candidate scheme and devise the WL-MC-SIC
MMSE algorithm which combines widely linear signal processing with
the MC-SIC MMSE scheme. Finally, the proposed interference
suppression strategy which employs the WL-MC-SIC MMSE scheme to
suppress the MAI and use the VSP scheme to suppress the
 {JS} is proposed.

\subsection{ VSP Scheme for Jamming Signal Subtraction}
Generally, the second-order statistics of the received signal $\boldsymbol{r}[i]$ are represented by the covariance matrix $\boldsymbol{R}_{rr}$ as
\begin{equation}
\begin{split}
\boldsymbol{R}_{rr} & = E[\boldsymbol{r}[i] \boldsymbol{r}[i]^{H}] = \boldsymbol{R}_{s} + \boldsymbol{R}_{j+n}, \label{cov_corr_matrix}
\end{split}
\end{equation}
where $\boldsymbol{R}_{s}$ is the covariance matrix of the desired signal and $\boldsymbol{R}_{j+n}$ denotes the covariance matrix without any contribution from the desired signal. In practical applications, the covariance matrix $ $can be estimated by using the time averaging method shown as follows
\begin{equation}
\begin{split}
{\hat{\boldsymbol{R}}}_{rr} & = \dfrac{1}{M}\sum_{i=1}^{M} \boldsymbol{r}[i] \boldsymbol{r}[i]^{H}, \label{cal_Rrr}
\end{split}
\end{equation}
where $M$ is the length of the averaging window.

Performing an eigendecomposition on $\boldsymbol{R}_{rr}$ yields
\begin{equation}
\begin{aligned}
{\boldsymbol{R}}_{rr} & = \sum_{i=1}^{L} \lambda_{i} \boldsymbol{e}_i (\boldsymbol{e}_{i})^{H}\\
&= \boldsymbol{E}_{s} \boldsymbol{D}_{s} (\boldsymbol{E}_{s})^{H} + \sigma^{2}_{n} \boldsymbol{E}_{n} (\boldsymbol{E}_{n})^{H},
\end{aligned}
\end{equation}
where $\{\lambda_{i},i=1,\cdots,L\}$ are the eigenvalues of $\boldsymbol{R}_{rr}$ arranged in decreasing order, $\boldsymbol{e}_{i}$ is the eigenvector associated with $\lambda_{i}$, $\boldsymbol{E}_{s}$ and $\boldsymbol{E}_{n}$ contain the $K+1$ dominant eigenvectors and the remaining eigenvectors, respectively. The quantity $K$ is the number of users, that is
\begin{equation}
\begin{split}
{\boldsymbol{E}}_{s} & = [\boldsymbol{e}_{1},\boldsymbol{e}_{2},\cdots,\boldsymbol{e}_{K+1}] \in \mathbb{C}^{L\times (K+1)},
\end{split}
\end{equation}
\begin{equation}
\begin{split}
{\boldsymbol{E}}_{n} & = [\boldsymbol{e}_{K+2},\boldsymbol{e}_{K+3},\cdots,\boldsymbol{e}_{L}] \in \mathbb{C}^{L\times (L-K-1)}
\end{split}
\end{equation}
and
\begin{equation}
\begin{split}
{\boldsymbol{D}}_{s} & = \textrm{diag}\{\lambda_{1},\lambda_{2},\cdots,\lambda_{K+1}\}
\end{split}
\end{equation}
is a diagonal matrix.

We can get the linear subspace $\mathcal{C}_{1} = \mathcal{L}[\boldsymbol{H}]$ where
\begin{equation}
\boldsymbol{H}=[\boldsymbol{s}_{1},\cdots,\boldsymbol{s}_{K}]
\end{equation}
and
\begin{equation}
\mathcal{C}_{2}=\mathcal{L}[\boldsymbol{E}_{s}].
\end{equation}
The desired user signal $\boldsymbol{u}_{k}=A_{k} b_{k}[i] \boldsymbol{s}_{k}[i]$, $\{k=1,\cdots,K \}$ lies in the subspace $\mathcal{C}_{1}$ and $\mathcal{C}_{2}$.
Therefore, we have the desired user signal lies within the intersection of $\mathcal{C}_{0}$ \cite{{Zhuang}} where
\begin{equation}
\mathcal{C}_{0}= \mathcal{C}_{1}\cap \mathcal{C}_{2}.
\end{equation}
In accordance with the theorem of sequential vector space projection \cite{Stark}, the intersection of the two constraint sets can be found by applying the alternating projection algorithm which is described by
\begin{equation}
\hat{\boldsymbol{U}}=\mathcal{P}\{\mathbb{P}_{\mathcal{C}_{2}} \mathbb{P}_{\mathcal{C}_{1}} \} \label{project_operator},
\end{equation}
where $\hat{\boldsymbol{U}}\in \mathbb{C}^{L \times K}$ is the estimate of the desired signal. $\mathcal{P}\{\boldsymbol{A}\}$ denotes the matrix composed of the $K$ eigenvectors of the matrix $\boldsymbol{A}$ and those $K$ dominant eigenvectors are selected according to the descending order of the eigenvalue of the matrix $\boldsymbol{A}$. $\mathbb{P}_{\mathcal{C}_{1}}$ and $\mathbb{P}_{\mathcal{C}_{2}}$ are the projection operators which are defined as \cite{{Zhuang}}
\begin{equation}
\mathbb{P}_{\mathcal{C}_{1}}=\boldsymbol{H}({\boldsymbol{H}}^{H} \boldsymbol{H})^{-1}\boldsymbol{H}^{H},
\end{equation}
\begin{equation}
\mathbb{P}_{\mathcal{C}_{2}}=\boldsymbol{E}_{s}(\boldsymbol{E}_{s})^{H}.
\end{equation}
Equation {(\ref{project_operator})} can be interpreted as selecting the vectors located in the linear subspace spanned by the columns of $\boldsymbol{H}$, which has the smallest angle from the subspace $\mathcal{L}[\boldsymbol{E}_{s}]$.

Using the estimated desired signal $\hat{\boldsymbol{U}}$, the desired-signal-absent covariance matrix can be formed by
\begin{equation}
\hat{\boldsymbol{R}}_{j+n}=\boldsymbol{R}_{rr}-\hat{\boldsymbol{U}}\hat{\boldsymbol{U}}^{H}. \label{desired-signal-absent}
\end{equation}
Performing an eigendecomposition on $\hat{\boldsymbol{R}}_{j+n}$ yields
\begin{equation}
\begin{aligned}
\hat{\boldsymbol{R}}_{j+n} &= \sum_{i=1}^{L} \hat{\lambda}_{i} \hat{\boldsymbol{e}}_i (\hat{\boldsymbol{e}}_{i})^{H}\\
&= \hat{\boldsymbol{E}}_{I}\hat{\boldsymbol{D}}_{I}\hat{\boldsymbol{E}}_{I}^{H}
+\hat{\boldsymbol{E}}_{N}\hat{\boldsymbol{D}}_{N}\hat{\boldsymbol{E}}_{N}^{H},\label{absent_signal_matrix}
\end{aligned}
\end{equation}
where $\{\hat{\lambda}_{i},i=1,\cdots,L\}$ are the eigenvalues of $\hat{\boldsymbol{R}}_{j+n}$ arranged in decreasing order, $\hat{\boldsymbol{e}}_{i}$ is the eigenvector associated with $\hat{\lambda}_{i}$. In addition,  $\hat{\boldsymbol{D}}_{I}$ and $\hat{\boldsymbol{D}}_{N}$ are diagonal matrices
and $\hat{\boldsymbol{E}}_{I}^{H}$ and $\hat{\boldsymbol{E}}_{N}^{H}$ consist of the $M$ dominant eigenvectors and remaining eigenvectors,
respectively. An important conclusion through simulations results shown below is drawn that the main power of the jamming interference
is centralized in the principal eigenvector of desired-signal-absent covariance matrix. Before we use the principal
eigenvector $\hat{\boldsymbol{e}}_{1}$, $\hat{\boldsymbol{e}}_{1}$ should be normalized as follows
\begin{equation}
\hat{\boldsymbol{e}}=\dfrac{\hat{\boldsymbol{e}}_{1}}{\sqrt{\langle \hat{\boldsymbol{e}}_{1},\hat{\boldsymbol{e}}_{1}} \rangle}
=\dfrac{\hat{\boldsymbol{e}}_{1}}{\sqrt{(\hat{\boldsymbol{e}}_{1}^{H} \hat{\boldsymbol{e}}_{1})}},
\end{equation}
where $\langle{\boldsymbol{a}},\boldsymbol{b} \rangle$ denotes the inner product of vector $\boldsymbol{a}$ and vector $\boldsymbol{b}$.

\begin{figure}[htb]
\begin{center}
\def\epsfsize#1#2{1\columnwidth}
\epsfbox{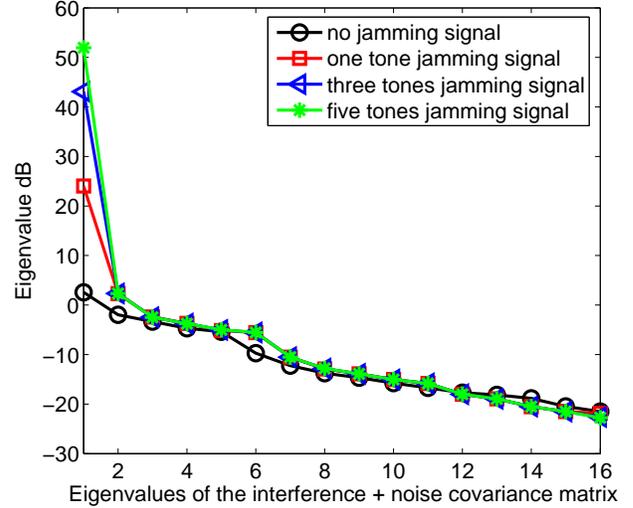} \vspace{-1em} \centering \caption{\footnotesize
The eigenvalue of the desired-signal-absent covariance matrix.}
\label{Fig1}
\end{center}
\end{figure}

Let $\mathbb{P}_{I}^{\perp}$ be the complement projection operator of the interference signal. Then $\mathbb{P}_{I}^{\perp}$ can be estimated by
\begin{equation}
\mathbb{P}_{I}^{\perp}=\boldsymbol{I}_{L}-\hat{\boldsymbol{e}}(\hat{\boldsymbol{e}})^H,
\end{equation}
where $\boldsymbol{I}_{L}$ is the $L\times{L}$ identity matrix. Using the complement projection operator $\mathbb{P}_{I}^{\perp}$ on the received signal vector $\boldsymbol{r}$ to suppress the JS before the SIC MMSE detection yields
\begin{equation}
\boldsymbol{r}^{\prime} = \mathbb{P}_{I}^{\perp} \boldsymbol{r},
\end{equation}
where $\boldsymbol{r}^{\prime}$ is the vector which is projected
onto the complement of the  {JS} subspace. The VSP algorithm to
preprocess the received signal in order to suppress the
 {JS} is summarized in Table
{\ref{table:VSP_algorithm}}.
\begin{table}[ht]
\caption{THE VSP ALGORITHM}
\centering
\begin{tabular}{l}
\hline\\[0.25ex]
1:~~Initialize the averaging windows $M$=100.~~~~~~~~~~~~~~~~~~~~~~~~~~~~~~~~~~~\\[0.5ex]
2:~~Calculate the covariance matrix \\
~~~~~~~~~~~~~~~~~${\hat{\boldsymbol{R}}}_{rr} =  \dfrac{1}{M}\sum_{i=1}^{M} \boldsymbol{r}[i] \boldsymbol{r}[i]^{H}$.\\[1ex]
3:~~Perform an eigendecomposition on $\hat{\boldsymbol{R}}_{rr}$ \\
~~~~~~~~~~~~~~~~~$\hat{{\boldsymbol{R}}}_{rr} = \boldsymbol{E}_{s} \boldsymbol{D}_{s} (\boldsymbol{E}_{s})^{H} + \sigma^{2}_{n} \boldsymbol{E}_{n} (\boldsymbol{E}_{n})^{H}$. \\[1ex]
4:~~~~~~~~~~~~~~~${\boldsymbol{E}}_{s} = [\boldsymbol{e}_{1},\boldsymbol{e}_{2},\cdots,\boldsymbol{e}_{K+1}]$.\\[1ex]
5:~~~~~~~~~~~~~~~$\boldsymbol{H} = [\boldsymbol{s}_{1},\cdots,\boldsymbol{s}_{K}]$.\\[1ex]
6:~~Calculate the projection operators\\
~~~~~~~~~~~~~~~~~$\mathbb{P}_{\mathcal{C}_{1}}=\boldsymbol{H}({\boldsymbol{H}}^{H} \boldsymbol{H})^{-1}\boldsymbol{H}^{H}$,\\[1ex]
~~~~~~~~~~~~~~~~~$\mathbb{P}_{\mathcal{C}_{2}}=\boldsymbol{E}_{s}(\boldsymbol{E}_{s})^{H}$.\\ [1ex]
7:~~~~~~~~~~~~~~~$\hat{\boldsymbol{U}}=\mathcal{P}\{\mathbb{P}_{\mathcal{C}_{2}} \mathbb{P}_{\mathcal{C}_{1}} \}$.\\[1ex]
8:~~Calculate the desired-signal-absent covariance matrix\\
~~~~~~~~~~~~~~~~~$\hat{\boldsymbol{R}}_{j+n}=\boldsymbol{R}_{rr}-\hat{\boldsymbol{U}}\hat{\boldsymbol{U}}^{H}$.\\[1ex]
9:~~Perform an eigendecomposition on $\hat{\boldsymbol{R}}_{j+n}$\\[1ex]
~~~~~~~~~~~~~~~~~$\hat{\boldsymbol{R}}_{j+n}= \sum_{i=1}^{L} \hat{\lambda}_{i} \hat{\boldsymbol{e}}_i (\hat{\boldsymbol{e}}_{i})^{H}$.\\[1ex]
10:~~~~~~~~~~~~~~$\hat{\boldsymbol{e}}=\dfrac{\hat{\boldsymbol{e}}_{1}}{\sqrt{(\hat{\boldsymbol{e}}_{1}^{H} \hat{\boldsymbol{e}}_{1})}}$.\\[1ex]
11:~~~~~~~~~~~~~~$\mathbb{P}_{I}^{\perp}=\boldsymbol{I}_{L}-\hat{\boldsymbol{e}}(\hat{\boldsymbol{e}})^H$.\\[1ex]
12:~~~~~~~~~~~~~~$\boldsymbol{r}^{\prime} = \mathbb{P}_{I}^{\perp} \boldsymbol{r}$.\\[1ex]
\hline
\end{tabular}
\label{table:VSP_algorithm} 
\end{table}
From the previous discussion we know that one important step for the
VSP scheme is detecting the existence of  {the jamming signal}. From
({\ref{desired-signal-absent}}) we can get the desired-signal-absent
covariance matrix $\hat{\boldsymbol{R}}_{j+n}$ and most of the
information about  {the jamming signal} is associated with the
principal eigenvector. This conclusion can be drawn from the
simulation result shown below. The simulation scenarios are: the
length of the spreading sequence is 16 and the spreading sequence is
randomly generated. The number of users is 8 and the signal to noise
ratio (SNR) for every user is set to 8dB. The power between
different tone interferences is equal and the signal to interference
ratio (SIR) of the tone interference is 0dB. The normalized
frequency of the tone interference is set as: $f_{1}=20$,
$f_{2}=40$, $f_{3}=60$, $f_{4}=80$, $f_{5}=100$. In this paper, SIR
is calculated after despreading. The eigenvalue of the
desired-signal-absent covariance matrix is shown in
Fig.{\ref{Fig1}}.

From the simulation result we can see that the interference energy
is mainly located at the principal eigenvalue of the
desired-signal-absent covariance matrix. Using this character we can
easily know about the existence of the  {JS}. This method is also
insensitive to the number of the interferers.
\subsection{ Widely Linear Signal Processing Scheme}
In \cite{Chevalier} and \cite{Schreier}, it has been shown that for BPSK modulation and other non-circular or improper modulation schemes, the performance of linear receivers can be further improved if both the received signal and also its complex conjugate are processed. This is because, for improper signals, the covariance matrix $\boldsymbol{R}_{rr}=E[\boldsymbol{r} \boldsymbol{r}^H]$ cannot completely describe the second-order statistics of the received vector and the complementary covariance matrix $\hat{\boldsymbol{R}}=E[\boldsymbol{r} \boldsymbol{r}^T]$ needs to be taken into account. The resulting receiver is referred to as a WL receiver.

In order to exploit the second-order information, we perform WL processing that utilizes the received vector $\boldsymbol{r}$ and its complex conjugate
$\boldsymbol{r}^{\ast}$ to form an augmented vector. For convenience we introduce the bijective transform $\mathcal{J}\{\cdot\}$
\begin{equation}
\boldsymbol{r}\to\tilde{\boldsymbol{r}}:~~~~~~~~\tilde{\boldsymbol{r}}=\dfrac{1}{\sqrt{2}}[\boldsymbol{r}^T,\boldsymbol{r}^{H}]^{T}.
\end{equation}
In what follows, all WL based quantities are denoted by an over tilde. An important property of $\mathcal{J}\{\cdot\}$ is that, for a complex vector
$\boldsymbol{r}$ and $\boldsymbol{u}$, $\tilde{\boldsymbol{u}}^{H}\tilde{\boldsymbol{r}}=\tilde{\boldsymbol{r}}^{H}\tilde{\boldsymbol{u}}$. Now a widely linear MMSE solution can be directly obtained as
\begin{equation}
\tilde{\boldsymbol{w}}_k=\boldsymbol{R}^{-1}_{\tilde{r}\tilde{r}}\tilde{\boldsymbol{p}}_{k},
\end{equation}
where
\begin{equation}
\boldsymbol{R}_{\tilde{r}\tilde{r}} = E[\tilde{\boldsymbol{r}}\tilde{\boldsymbol{r}}^{H}]
=\dfrac{1}{2}
\left[\begin{array}{cc} {\boldsymbol{R}}_{rr} & {\hat{\boldsymbol{R}}_{rr}} \\
{\hat{\boldsymbol{R}}_{rr}^{\ast}} & {\boldsymbol{R}}_{rr}^{\ast} \end{array}\right],
\end{equation}
\begin{equation}
\tilde{\boldsymbol{p}}_{k}=\mathcal{J}\left\lbrace {\boldsymbol{p}}_{k}\right\rbrace
\end{equation}
and the widely linear MMSE estimate of the data symbol is given by
\begin{equation}
\hat{b}_k=\mathcal{R}\{\textrm{sgn}(\tilde{\boldsymbol{w}}_k\tilde{\boldsymbol{r}}) \}.
\end{equation}

\subsection{MC-SIC MMSE Design}
As mentioned above, the SIC scheme is a decision-driven detection algorithm which suffers from error propagation and performance degradation, the strategy of the MC-SIC scheme is to find the optimum feedback decision and mitigate the error propagation. The key idea behind the MC-SIC scheme is to exploit the structural properties of the sub-constellation formed by the signals of two consecutive users in an ordered set to gain an improvement in detection performance.

In the following, we detail the MC-SIC algorithm through the procedure for detecting $\hat{\boldsymbol{s}}_{k}[i]$ for user $k$. The detection of the data symbols of the other user can be performed accordingly. The soft estimation of the user $k$'s symbol $u_k[i]$ is obtained by using the MMSE detector as
\begin{equation}
u_{k}[i]=\boldsymbol{w}_{k}^{H}\hat{\boldsymbol{r}}_{k}[i],
\end{equation}
where the MMSE filter is given by
$\boldsymbol{w}_{k}=(\boldsymbol{P}_{k}\boldsymbol{P}_{k}^{H}+\delta^{2}_{n}\boldsymbol{I}_{L})^{-1}\boldsymbol{p}_{k}$,
$\boldsymbol{P}_{k}$ denotes the matrix obtained by taking the columns $k,k+1,\cdots,K$ of $\boldsymbol{P}$ and $\hat{\boldsymbol{r}}_k[i]$ is the received vector after cancellation of the $k-1$ previously detected symbols.

\begin{figure}[htb]
\begin{center}
\def\epsfsize#1#2{0.90\columnwidth}
\epsfbox{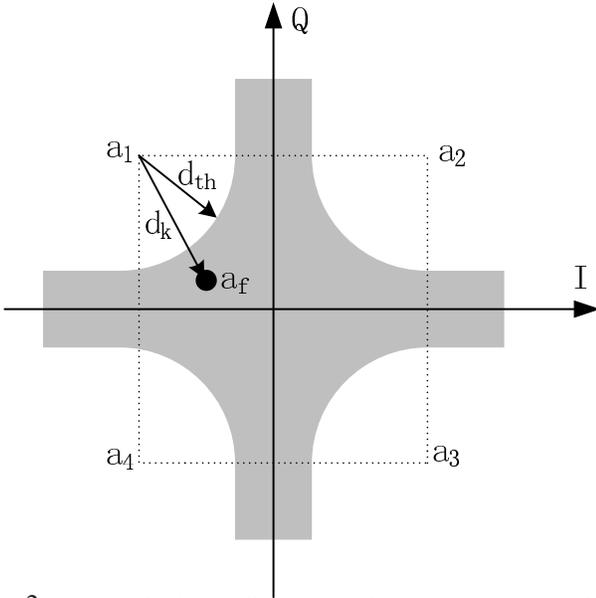} \vspace{-1em} \centering \caption{\footnotesize
The combined constellation constraint structure. The constellation
constraint is invoked as the soft estimate parameter $a_{f}$ is
dropped into the shaded area.} \label{Fig2}
\end{center}
\end{figure}

Based on the estimate of the user $k$’s symbol $u_{k}[i]$, the user $(k+1)$’s symbol can be calculated as
\begin{equation}
u_{k}[i]=\boldsymbol{w}_{k+1}^H(\hat{\boldsymbol{r}}_{k}[i]-Q[u_{k[i]}]\boldsymbol{p}_{k}),
\end{equation}
where $Q[\cdot]$ is the signal quantization operator used to detect the signals of each user. The detection parameter $a_{f}$ is constructed with the estimated symbols $u_k[i]$ and $u_{k+1}[i]$ as
\begin{equation}
a_{f} = \mathcal{R}\{u_k[i]\} + \textrm{j}\mathcal{R}\{u_{k+1}[i]\}.
\end{equation}
For each user, the reliability of the soft estimate $u_k[i]$ is determined by the combined constellation constraint structure which is illustrated in Fig.{\ref{Fig2}}.

The combined constellation constraint structure shown in Fig.{\ref{Fig2}} is for BPSK modulation mode and the combined constellation set is constructed
as $\mathcal{A}=\mathcal\{a_1=-1+\textrm{j}, a_2=1+\textrm{j}, a_3=1-\textrm{j}, a_4=-1-\textrm{j}\}$.
The parameter $d_{th}$ is the threshold distance to evaluate the reliability of the current estimated symbol $u_k[i]$, which is a predefined parameter.
The reliability of the estimated symbol is determined by the Euclidean distance between the detection parameter $a_f$ and its nearest combined
constellation points, which is given by
\begin{equation}
d_k=\mid {a}_{opt}-a_f\mid.
\end{equation}
The optimum parameter $a_{opt}$ denotes the constellation point which is the nearest to the detection parameter $a_f$ and can be expressed as
\begin{equation}
a_{opt}=\textrm{arg min}_{{a_{c}}\in \mathcal{A}}\{\mid a_c-a_f \mid\}.
\end{equation}
There are two possibilities as follows:

\begin{enumerate}
\item  If $d_{k}\geq{d_{th}}$, then the current soft estimate $u_k[i]$ is determined reliable and the estimated symbol of user $k$ is obtained by
$\hat{s}_k[i]=Q[u_k[i]]$. After we get the estimated symbol $\hat{s}_k[i]$, we can regenerate user $k$, cancel it from the received vector $\hat{\boldsymbol{r}}_k[i]$ and continue the procedure above to estimate user $k+1$.
\item  If $d_{k}<d_{th}$, then the current soft estimate $u_k[i]$ is determined unreliable and the optimum feedback symbol must be found before cancellation. Since the effect of “closest” interferer is significant in terms of performance degradation while estimating the user $k$’s symbol $\hat{s}_k[i]$ \cite{Singer}, we estimate two consecutive user signals $\{ s_k, s_{k+1}\}$ at one time and the candidates for $s_k$ and $s_{k+1}$ are selected
from $\mathcal{L}=\{\boldsymbol{c}_1,\boldsymbol{c}_2, \boldsymbol{c}_3,\boldsymbol{c}_4 \}$,
where $\boldsymbol{c}_1=[-1,+1]$, $\boldsymbol{c}_2=[-1,-1]$, $\boldsymbol{c}_3=[+1,+1]$, $\boldsymbol{c}_4=[+1,-1]$. After the optimum candidate $\boldsymbol{c}_{opt}$ is selected from $\mathcal{L}$, the effects of user $k$ and $k+1$ will be subtracted together before detecting the rest of the users.
The selection algorithm is described as follows.
\end{enumerate}

In order to get the optimum candidate $\boldsymbol{c}_{opt}$ for user $k$ and user $k+1$, we construct the estimated symbol vector from three parts:
the first part is the previously detected symbols $\hat{s}_1[i]$, $\hat{s}_2[i]$, $\cdots$, $\hat{s}_{k-1}[i]$, the second part $\boldsymbol{c}_k$
is the symbol taken from the candidate constellation point set $\mathcal{L}$, the last part uses the previous decisions and performs the following
users $k+2$,$\cdots$,$K$’s detection by the nulling and symbol cancellation which is equivalent to the traditional SIC algorithm. Therefore, we can
get the estimated symbol vector
\begin{equation}
\boldsymbol{b}^m[i]=[\hat{s}_1[i],\cdots,\hat{s}_{k-1}[i],\boldsymbol{c}_m,b^m_{k+2}[i],\cdots,b^m_q[i],\cdots,b^m_K[i]],
\end{equation}
where $\boldsymbol{c}_m \in \mathcal{L}$, $b^m_q[i]$ is the potential decision that corresponds to the selection of $\boldsymbol{c}_m$ in the constellation point set,
\begin{equation}
b^m_q[i]=Q\left[\boldsymbol{w}^m_k \hat{\boldsymbol{r}}^m_q[i]\right].
\end{equation}
where $q$ indexes a certain user between $k+2$ to $K$. For each user the same MMSE filter $\boldsymbol{w}_k$ is used for all the candidates, which can be calculated in advance and allows the proposed algorithm to have the computational simplicity of the SIC algorithm described by
\begin{equation}
\hat{\boldsymbol{r}}^m_q[i]=\hat{\boldsymbol{r}}^m_k[i]-\left[\boldsymbol{p}_k,\boldsymbol{p}_{k+1}\right]\boldsymbol{c}^T_m
-\sum_{j=k+2}^{q-1}\boldsymbol{p}_jb^m_j[i].
\end{equation}
According to the maximum likelihood rule in the selected candidates set, the optimum candidate $\boldsymbol{c}_{opt}$ is given by
\begin{equation}
\boldsymbol{c}_{opt}=\textrm{arg min}_{\boldsymbol{c}_m \in \mathcal{L}}{\parallel \boldsymbol{r}[i]-\boldsymbol{P}\boldsymbol{b}^m[i]\parallel }^2,
\end{equation}
where the $\boldsymbol{c}_{opt}$ is chosen to replace the unreliable $u_k[i]$ and $u_{k+1}[i]$, which will be the optimal feedback symbols for the
next user as well as the more reliable estimate for the current two users.

\subsection{Widely Linear MC-SIC MMSE Algorithm  }
In this subsection, we will combine widely linear signal processing with the MC-SIC MMSE scheme and obtain the WL-MC-SIC MMSE algorithm.

From the previous subsection, we can compute the widely linear MMSE filter as
\begin{equation}
\hat{\boldsymbol{w}}_k=2
\left(\begin{array}{cc} \left(\boldsymbol{P}_k \boldsymbol{P}^{H}_k + \sigma^2_{n}\boldsymbol{I}_L\right)
& \left( \boldsymbol{P}_k \boldsymbol{P}^T_k  \right)\\
\left( \boldsymbol{P}_k \boldsymbol{P}^T_k  \right)^{\ast}
& {\left(\boldsymbol{P}_k \boldsymbol{P}^{H}_k + \sigma^2_{n}\boldsymbol{I}_L\right)}^{\ast} \end{array}\right)^{-1} \tilde{\boldsymbol{p}}_{k}.
\end{equation}
The soft estimate of the user $k$’s symbol $\tilde{u}_k[i]$ is obtained by using the WL MMSE detector as
\begin{equation}
\tilde{u}_k[i]=\tilde{\boldsymbol{w}}_k^H \tilde{\boldsymbol{r}}_k[i],
\end{equation}
where $\tilde{r}_k[i]=\mathcal{J}\left\lbrace {\tilde{\boldsymbol{r}}}_k[i]\right\rbrace$.
The algorithm of the proposed WL-MC-SIC is summarized in Table {\ref{table:WL-MC_SIC}}.

\begin{table}[ht]
\caption{THE WL MC-SIC ALGORITHM}
\centering
\begin{tabular}{l}
\hline\\[0.25ex]
1:~~$\hat{\boldsymbol{w}}_k=2
\left(\begin{array}{cc} \left(\boldsymbol{P}_k \boldsymbol{P}^{H}_k + \sigma^2_{n}\boldsymbol{I}_L\right)
& \left( \boldsymbol{P}_k \boldsymbol{P}^T_k  \right)\\
\left( \boldsymbol{P}_k \boldsymbol{P}^T_k  \right)^{\ast}
& {\left(\boldsymbol{P}_k \boldsymbol{P}^{H}_k + \sigma^2_{n}\boldsymbol{I}_L\right)}^{\ast} \end{array}\right)^{-1}$~~~~~~~~~~~\\[3ex]
2:~~$\mathcal{L}=\left\lbrace \boldsymbol{c}_1, \boldsymbol{c}_2, \boldsymbol{c}_3, \boldsymbol{c}_4 \right\rbrace $\\[1ex]
3:~~ while $k<K$ \\
4:~~~~~$\tilde{u}_k[i] = \tilde{\boldsymbol{w}}_k^H \tilde{\boldsymbol{r}}_k[i]$\\[1ex]
5:~~~~~$\tilde{u}_{k+1}[i] = \tilde{\boldsymbol{w}}_{k+1}^H(\tilde{\boldsymbol{r}}_k[i]-Q\left[\tilde{u}_k[i]\right] \tilde{p}_k)$\\[1ex]
6:~~~~~$a_f = \mathcal{R}\left\lbrace u_k[i]\right\rbrace +\textrm{j}\mathcal{R}\left\lbrace u_{k+1}[i]\right\rbrace $\\[1ex]
7:~~~~~if $d_k>d_{th}$\\[1ex]
8:~~~~~~~if $k=K-1$\\[1ex]
9:~~~~~~~~~$\boldsymbol{c}_{opt}=\textrm{arg min}_{\boldsymbol{c}_m \in \mathcal{L}}{\parallel \boldsymbol{r}[i]
-\boldsymbol{P}\boldsymbol{b}^m[i]\parallel }^2$\\[1ex]
10:~~~~~~~~$ \tilde{s}_{K-1}[i]=\boldsymbol{c}_{opt}(1,1)~~~\tilde{s}_K[i]=\boldsymbol{c}_{opt}(1,2) $\\[1ex]
11:~~~~~~~~$\textrm{loopsym}=1$ ~~\% loopsym=1 denotes program execution in\\
~~~~~~~~~~~~~~~~~~~~~~~~~~~~~~~~~~this branch,initial value is 0\\
12:~~~~~~else if $k=K-2$\\[1ex]
13:~~~~~~~~~~~~~$\tilde{\boldsymbol{r}}^m_K[i]-[\tilde{\boldsymbol{p}}_{K-2},\tilde{\boldsymbol{p}}_{K-1}]\boldsymbol{c}^T_m $\\[1ex]
14:~~~~~~~~~~~~~$b^m_K[i]=Q[\tilde{\boldsymbol{w}}^H_K \tilde{\boldsymbol{r}}^m_K[i]] $\\[1ex]
15:~~~~~~~~~~~~~$\boldsymbol{b}^m[i]=\left[\tilde{s}_1[i],\cdots,\tilde{s}_{K-3}[i],\boldsymbol{c}_m,b^m_K[i] \right] $\\[1ex]
16:~~~~~~~~~~~~~$\boldsymbol{c}_{opt}=\textrm{arg min}_{\boldsymbol{c}_m \in \mathcal{L}}
\parallel \tilde{\boldsymbol{r}}[i]-\tilde{\boldsymbol{P}}\boldsymbol{b}^m[i]\parallel^2$\\[1ex]
17:~~~~~~~~~~~~~$\tilde{_{K-2}[i]}=\boldsymbol{c}_{opt}(1,1)~~\tilde{_{K-1}[i]}=\boldsymbol{c}_{opt}(1,2)$\\[1ex]
18:~~~~~~~~~~~~~$\tilde{\boldsymbol{r}}_k[i]=\tilde{\boldsymbol{r}}_{K-2}[i]-\tilde{s}_{K-2}[i]\tilde{\boldsymbol{p}}_{K-2}
-\tilde{s}_{K-1}[i]\tilde{\boldsymbol{p}}_{K-1}$\\
19:~~~~~~~~~~else\\
20:~~~~~~~~~~~~$\tilde{\boldsymbol{r}}_{k+2}^m[i]=\tilde{\boldsymbol{r}}_{k}^m[i]-\left[\tilde{\boldsymbol{p}}_{k-2},\tilde{\boldsymbol{p}}_{k-1}\right]
\boldsymbol{c}^T_m $\\[1ex]
21:~~~~~~~~~~~~for $j=k+2$ to $K$\\[1ex]
22:~~~~~~~~~~~~~~$b^m_j[i]=Q\left[\tilde{\boldsymbol{w}}^H_j \tilde{\boldsymbol{r}}^m_j[i]\right] $\\[1ex]
23:~~~~~~~~~~~~~~$\tilde{\boldsymbol{r}}^m_{j+1}=\tilde{\boldsymbol{r}}^m_{j}-b^m_j[i]\tilde{\boldsymbol{p}}_j$\\[1ex]
24:~~~~~~~~~~~~end for\\
25:~~~~~~~~~~~~$\boldsymbol{c}_{opt}=\textrm{arg min}_{\boldsymbol{c}_m \in \mathcal{L}}
\parallel \tilde{\boldsymbol{r}}[i]-\tilde{\boldsymbol{P}}\boldsymbol{b}^m[i]\parallel^2$\\[1ex]
26:~~~~~~~~~~~~$ \tilde{s}_{k}[i]=\boldsymbol{c}_{opt}(1,1)~~~\tilde{s}_{k+1}[i]=\boldsymbol{c}_{opt}(1,2) $\\[1ex]
27:~~~~~~~~~~~~$\tilde{\boldsymbol{r}}_{k+2}[i]=\tilde{\boldsymbol{r}}_{k}[i]-\tilde{s}_{k}[i]\tilde{\boldsymbol{p}}_{k}
-\tilde{s}_{k+1}[i]\tilde{\boldsymbol{p}}_{k+1}$\\[1ex]
28:~~~~~~end if\\
29:~~~~~~$k=k+2$\\[1ex]
30:~~~else\\
31:~~~~~$\tilde{s}_k[i]=Q\left[\tilde{u}_k[i] \right]$\\[1ex]
32:~~~~~$\tilde{\boldsymbol{r}}_{k+1}[i]=\tilde{\boldsymbol{r}}_k[i]-\tilde{s}_k[i] \tilde{\boldsymbol{p}}_k$\\[1ex]
33:~~~~~$k=k+1$\\[1ex]
34:~~~end if\\
35:~~end while\\
36:~~if $\textrm{loopsym}\not=1$~~~~~~~\\[1ex]
37:~~~~~$\tilde{s}_K[i]=Q\left[\tilde{\boldsymbol{w}}_{K}^H \tilde{\boldsymbol{r}}_{K}[i] \right]  $\\[1ex]
38:~~end if\\[1ex]
\hline
\end{tabular}
\label{table:WL-MC_SIC} 
\end{table}

\subsection{VSP Scheme and WL-MC-SIC MMSE Detector}
In this subsection, the procedure for MAI and  {JS} suppression
using the proposed WL-MC-SIC MMSE algorithm along with the VSP
scheme is described.

Using the VSP algorithm \cite{{Zhuang}} we can get the desired-signal-absent covariance matrix $\hat{\boldsymbol{R}}_{j+n}$ and perform an eigen-decomposition as in (\ref{absent_signal_matrix}). The eigenvalues $\{ \hat{\lambda}_i,~i=1,\cdots,L \}$ are obtained and we can set a threshold $d_{\lambda}$ for checking the existence of the JS, which can be calculated as follows
\begin{equation}
d_{\lambda}=\beta\left( \dfrac{1}{D} \sum_{i=1}^{D}\hat{\lambda}_{i+1}\right),
\end{equation}
where $\beta$ is a threshold factor and its associated with the
power of the  {JS} and the ambient noise. In this paper it is set to
$0.2$. $D$ is the number of eigenvalues which are used to estimate
the power of the ambient noise and normally it can be set as $D=2$.
By comparing the principal eigenvalue $\hat{\lambda}_1$ with the
threshold $d_{\lambda}$, we can get the information about the
existence of  {the jamming signal}. The procedure for MAI and
 {JS} suppression using the proposed algorithm is
summarized in Table {\ref{table:Interference suppression for MAI and
JS}}.
\begin{table}[ht]
\caption{THE ANTI-INTERFERENCE PROCEDURE of VSP and WL-MC-SIC MMSE ALGORITHM}
\centering
\begin{tabular}{l}
\hline\\[0.25ex]
Step1: Calculate the covariance matrix $\hat{\boldsymbol{R}}_{rr}$.~~~~~~~~~~~~~~~~~~~~~~~~~~~~~~~~~\\[1ex]
Step2: Calculate the desired-signal-absent covariance matrix $\hat{\boldsymbol{R}}_{j+n}$\\
~~~~~~~~using VSP scheme.\\[1ex]
Step3: Perform an eigendecomposition on $\hat{\boldsymbol{R}}_{j+n}$ \\
~~~~~~~~and get the eigenvalues $\{ \hat{\lambda}_i,~i=1,\cdots,L\}$.\\[1ex]
Step4: Calculate the threshold $d_{\lambda}$.\\[1ex]
Step5: Compare the principal eigenvalue $\hat{\lambda}_1$ with the threshold $d_{\lambda}$,\\
~~~~~~~~if $\hat{\lambda}_1>d_{\lambda}$, execute step 6 otherwise execute step 7.\\[1ex]
Step6: Perform the  {JS} suppression using the VSP algorithm.\\[1ex]
Step7: Perform MAI suppression using the WL-MC-SIC MMSE\\
~~~~~~~~and get the estimate of the desired signal.\\[1ex]
\hline
\end{tabular}
\label{table:Interference suppression for MAI and JS} 
\end{table}

\subsection{Complexity Analysis}
In this section, we will discuss the computational complexity of the proposed WL-MC-SIC MMSE algorithm compared with other existing algorithms mentioned above.

The $k$-th user is the desired user in this paper, and we should deduct the effects of $k$-1 users before we estimate the $k$-th user. In terms of complex multiplications, the complexity of the proposed algorithm and other existing algorithms as mentioned above is presented in Table {\ref{table:complexity}}. We focus only on the complexity of the main successive cancellation procedure since the rest of the operations including the VSP procedure are similar to the algorithms to be compared against.

\begin{table}[ht]
\caption{THE ANTI-INTERFERENCE PROCEDURE OF VSP AND WL-MC-SIC MMSE ALGORITHM}
\centering
\begin{tabular}{|l|c|}
\hline
Algorithm & ~~~~~Required complex multiplications~~~~~~~\\
\hline
$\textrm{SIC MMSE}$      & $RN$\\[1ex]
\hline
$\textrm{MF-SIC MMSE}$   & $RN+MRSN+QRN$\\[1ex]
\hline
$\textrm{MC-SIC MMSE}$   & $RN+MRS^2N+PRN$\\[1ex]
\hline
$\textrm{WL-SIC MMSE}$   & $2RN$\\[1ex]
\hline
$\textrm{WL-MF-SIC MMSE}$ & $RN+2MRSN+2QRN$\\[1ex]
\hline
$\textrm{WL-MC-SIC MMSE}$ & $RN+2MRS^2N+2PRN$\\[1ex]
\hline
\end{tabular}
\label{table:complexity} 
\end{table}

The main difference between the SIC MMSE algorithm and the MC-SIC MMSE detector is how to choose the optimum candidates from
the constellation points to replace the unreliable estimate. The threshold $d_{th}$ is an important factor on the effect
of the algorithm complexity. In Table {\ref{table:complexity}} the parameter $R$ denotes the number of times that the
optimum candidate will be calculated when the threshold $d_{th}$ is set. The parameter $N$ denotes the length of the MMSE filter.
The parameter $S$ denotes the number of candidates in $\mathcal{L}$. The parameter $M$ denotes the unreliable estimate times.
The parameters $Q$ and $P$ denote the number of times we need to calculate the third part of the estimated
symbol vector $\boldsymbol{b}^m[i]$ in the MF-SIC and MC-SIC algorithms respectively and usually $P$ is bigger than $Q$.

The parameter $d_{th}$ has an important influence on the performance and the complexity of algorithms. we should find a trade off between them. In the next section we will simulate the effects of the different $d_{th}$ on the algorithm.

\section{SIMULATION RESULTS}
In this section, we assess the bit error rate (BER) performance of the proposed WL-MC-SIC MMSE algorithm compared with the existing MAI cancellation algorithms mentioned above. Firstly, the simulations for the selection of parameter threshold $d_{th}$ are made. Secondly, the performance comparisons
between the proposed WL-MC-SIC MMSE algorithm and the existing MAI cancellation algorithms are made. Finally, the performance of the MAI and JS
suppression of the proposed VSP and WL-MC-SIC MMSE is shown.

In the following simulations, we consider that all the algorithms are used in synchronous DS-CDMA systems employing BPSK modulation and we transmit 10000 information symbols per user in one packet and the results are averaged over independent Monte Carlo runs. The spreading signature used for every user is randomly generated and the sequence length is 16. We assume that all the users have the same power unless otherwise stated.

\subsection{ Parameter Threshold Setting}
\begin{figure}[htb]
\begin{center}
\def\epsfsize#1#2{1\columnwidth}
\epsfbox{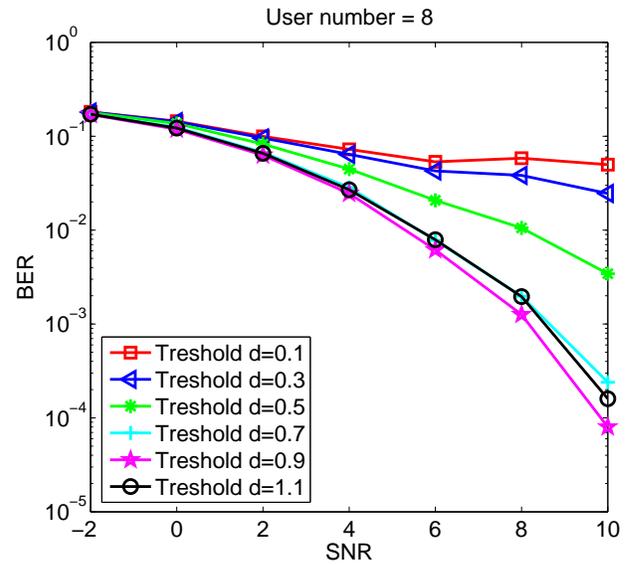} \vspace{-1em} \centering \caption{\footnotesize
BER performance against SNR(dB) with threshold $d_{th}$ = 0.1, 0.3,
0.5, 0.7, 0.9 and 1.1.} \label{Fig3}
\end{center}
\end{figure}

As we have discussed above, the threshold $d_{th}$ is crucial to the performance of the proposed algorithm. The threshold $d_{th}$ can be either a constant or a function of the signal power and the noise power, which can also be obtained through simulation in practical applications. In the paper we assess the BER performance of the algorithm under different thresholds to get the optimum threshold. We assume that the number of users is 8 and the SNR varies from -2dB to 12dB. The threshold is set to $d_{th}$ = 0.1, 0.3, 0.5, 0.7, 0.9 and 1.1. The evaluation of the BER performance against the SNR with different values of $d_{th}$ is shown in Fig.{\ref{Fig3}}.

\begin{figure}[htb]
\begin{center}
\def\epsfsize#1#2{1\columnwidth}
\epsfbox{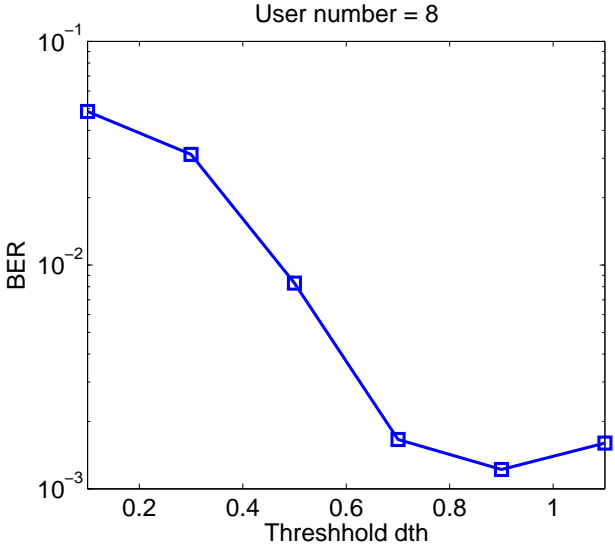} \vspace{-1em} \centering \caption{\footnotesize
BER performance against threshold with $\textrm{SNR}=8\textrm{dB}$
and the number of user is set to 8.} \label{Fig4}
\end{center}
\end{figure}

In order to detail the performance difference between the different values of the threshold $d_{th}$, we assess the BER performance of the algorithm at SNR=8dB. The simulation result is shown in Fig.{\ref{Fig4}}.

From the simulation results shown in Fig.{\ref{Fig3}} and Fig.{\ref{Fig4}} we can get the optimum threshold $d_{th}=0.9$. It should be pointed out
that the optimum value of $d_{th}$ is relative and it also varies with the different scenarios. The optimum value should be selected under the different scenarios. Unless stated otherwise, in the following scenarios in this paper the threshold $d_{th}$ of the proposed WL-MC-SIC MMSE
algorithm is set to $d_{th}=0.9$.

\subsection{ Performance Comparison }
In this subsection, we will compare the BER performance of the proposed WL-MC-SIC MMSE algorithm with the existing MAI cancellation algorithms mentioned above. We use SNR, the capacity in terms of the number of users and algorithm complexity to compare the performance of those algorithms.

Firstly, we evaluate the BER performance against the SNR of the received signal, for the MF-SIC algorithm we choose the optimum threshold $d_{th}=0.3$
which is also obtained through simulation in the same simulation scenario. In the following simulations, we choose this value for the MF-SIC algorithm
and the WL-MF-SIC algorithm. The number of users in this scenario is set to 8.

\begin{figure}[htb]
\begin{center}
\def\epsfsize#1#2{1\columnwidth}
\epsfbox{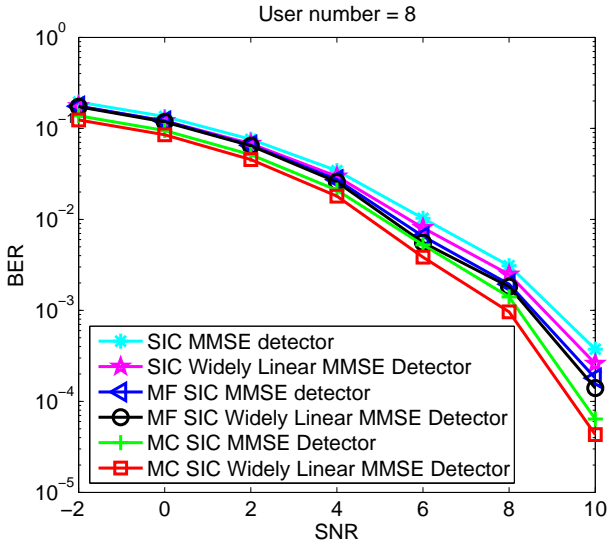} \vspace{-1em} \centering \caption{\footnotesize
BER performance against SNR with the number of users is set to 8 and
SNR varies from -2dB to 10dB.} \label{Fig5}
\end{center}
\end{figure}

The simulation result in Fig.{\ref{Fig5}} shows that the BER performance of the WL-MC-SIC MMSE algorithm outperforms other algorithms.

\begin{figure}[htb]
\begin{center}
\def\epsfsize#1#2{1\columnwidth}
\epsfbox{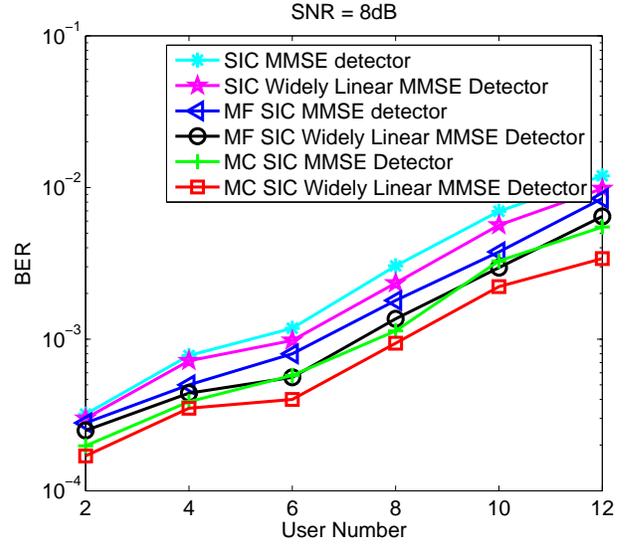} \vspace{-1em} \centering \caption{\footnotesize
BER performance against the number of users with SNR=8dB and the
number of user varies from 2 to 12.} \label{Fig6}
\end{center}
\end{figure}

Secondly, we evaluate the BER performance against the number of users. The SNR of this scenario is set to 8dB. The simulation result is illustrated
in Fig.{\ref{Fig6}}. From the simulation result shown in Fig.{\ref{Fig6}} we can see that with the increase in the number of users,
the proposed algorithm has a better performance than the other considered algorithms.

\begin{figure}[htb]
\begin{center}
\def\epsfsize#1#2{1.0\columnwidth}
\epsfbox{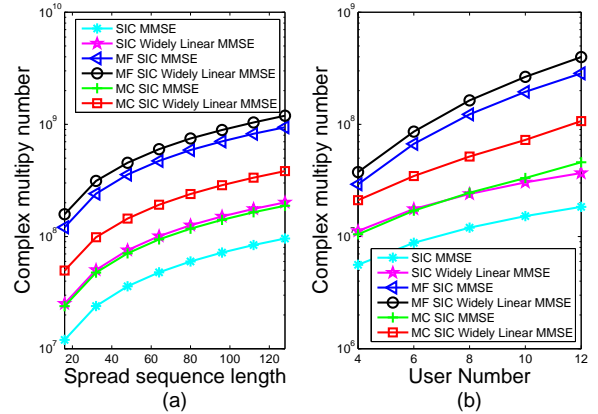} \vspace{-1em} \centering \caption{\footnotesize
Algorithm complexity simulation, (a) algorithm complexity against
the length of spreading sequence, SNR=8dB and the number of users is
set to 8, (b)algorithm complexity against the number of users,
SNR=8dB and the length of spreading sequence is set to 16.}
\label{Fig7}
\end{center}
\end{figure}

Finally, we will evaluate the algorithm complexity of those algorithms. We have analyzed the algorithm complexity theoretically
in Section IV. Here two simulation scenarios will be considered: the first scenario is the algorithm complexity against
the length of spreading sequence. The number of users is set to 8 and SNR is set to 8dB. The length of spreading sequence varies from 16 to 128.
As mentioned above, we use the number of complex multiplications as the tool to evaluate the complexity of those algorithms.
Fig.{\ref{Fig7}}(a) shows the simulation result of the first scenario.
the second scenario is the algorithm complexity against the number of users and the simulation is shown in Fig.{\ref{Fig7}}(b). In the simulation
SNR is set to 8dB and the number of users varies from 4 to 12.

From the simulation results shown in Fig.{\ref{Fig7}}, we find the WL-MC-SIC MMSE algorithm has much less complexity than the WL-MF-SIC MMSE algorithm and the MF-SIC MMSE algorithm. Since the proposed algorithm uses the candidate scheme to alleviate the effect of error
propagation in the conventional SIC algorithm, the complexity of the proposed algorithm is inevitably higher than the conventional SIC algorithm.

\subsection{Performance of Jamming Suppression }
The linear MMSE detection algorithm has a good performance for
suppressing the MAI and the JS \cite{Poor1}.  In order to get the
MMSE detector, we should have the knowledge of the jamming signal,
usually which can be obtained through an adaptive strategy
\cite{Poor2}. In this subsection we evaluate the performance of the
WL-MC-SIC algorithm on joint MAI and  {JS} suppression.

We take the tone interference as  {the jamming signal} and its form
is shown in formula ({\ref{jamming_signal}}). From the conclusion we
have drawn above we know that the VSP algorithm is insensitive to
the number of tone interferers. In the simulation below, we consider
a scenario that uses only one tone signal as the
 {JS} and the parameter $m=1$. As the VSP algorithm
is used to suppress the  {JS} before suppressing the MAI, to be
simplicity we only assess the effect of  {JS} suppression of the
proposed algorithm, which combines the VSP algorithm with the WL-MC
MMSE algorithm together. The other simulation conditions are the
same with the simulations above.

\begin{figure}[htb]
\begin{center}
\def\epsfsize#1#2{1\columnwidth}
\epsfbox{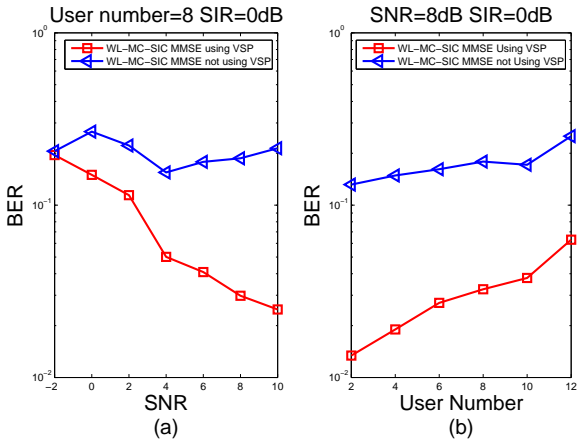} \vspace{-1em} \centering \caption{\footnotesize
BER performance simulations under the   {JS} scenario and SIR=0dB:
(a) BER performance against SNR, the number of users is set to 8,
(b) BER performance against the number of user, SNR=8dB.}
\label{Fig9}
\end{center}
\end{figure}

From the simulation results shown in Fig.{\ref{Fig9}} we can see
that using the VSP algorithm to suppress the   {JS} before detecting
the desired signals will improve the performance of the SIC detector
greatly. The proposed algorithm which combines the VSP method and
WL-MC-SIC MMSE algorithm together has a much better performance for
MAI and   {JS} suppression than the existing schemes mentioned
above.

\section{Conclusion}

A WL-MC-SIC MMSE algorithm for DS-CDMA systems for joint MAI and
{JS} suppression is proposed in this paper. The contributions of our
research work are composed of three parts: Firstly, in order to
cancel  {the jamming signal}, we present a VSP algorithm which is
effective in the  {JS} suppression. This method is able to identify
the existence of  {the jamming signal} and is insensitive to the
number of the tones. Secondly, we propose the multiple candidates
constellation constraints scheme to alleviate the effect of error
propagation in the SIC algorithm. In order to deal with the improper
characteristic of received signal, widely-linear signal processing
is used to make full use of the second-order information of received
vector which outperforms linear signal processing. Finally, we
propose a WL-MC-SIC MMSE scheme which combines the VSP algorithm
with the WL-MC-SIC MMSE algorithm together for joint suppressing MAI
and  {JS}. The simulation results show that the proposed algorithm
outperforms the existing SIC MMSE algorithms mentioned above and
have a good performance for  {JS} suppression.

\end{document}